\documentclass[aps,prb,a4paper,onecolumn,showpacs,showkeys,fleqn,floatfix,unsortedaddress]{revtex4-1}

\usepackage[dvips]{epsfig}
\usepackage[dvips]{color}
\usepackage{amsmath}

\newcommand{\figwidth}{0.9\columnwidth}
\newcommand{\sz}{\widehat{S}_{Z_A}{}}
\newcommand{\sx}{\widehat{S}_{X_A}{}}
\newcommand{\sy}{\widehat{S}_{Y_A}{}}

\newcommand{\svect}{\widehat{\mathbf{S}}}
\newcommand{\vect}[1]{\mathbf{#1}}
\newcommand{\ham}{\widehat{\cal{H}}}

\newcommand{\boldeta}{\boldsymbol\eta}

\begin{document}
\title{Electron spin resonance study of anisotropic
interactions in a two-dimensional spin gap magnet PHCC}

\author{V.N. Glazkov}
\email{glazkov@kapitza.ras.ru}

\affiliation{Kapitza Institute for Physical Problems RAS, Kosygin
str. 2, 119334 Moscow, Russia}

\affiliation{Neutron Scattering and Magnetism, Institute for Solid
State Physics, ETH Zurich, Switzerland,  8093 Z\"{u}rich,
Switzerland}

\author{T.S. Yankova}
\affiliation{Neutron Scattering and Magnetism, Institute for Solid
State Physics, ETH Zurich, Switzerland,  8093 Z\"{u}rich,
Switzerland}

\altaffiliation[Permanent address: ]{ Chemical Department,
M.V.Lomonosov Moscow State University, Moscow, Russia}

\author{J. Sichelschmidt}

\affiliation{ Max Planck Institute for Chemical Physics of Solids,
N\"{o}thnitzer Stra{\ss}e 40, 01187 Dresden, Germany}

\author{D. H\"{u}vonen}

\affiliation{Neutron Scattering and Magnetism, Institute for Solid
State Physics, ETH Zurich, Switzerland,  8093 Z\"{u}rich,
Switzerland}

\author{A. Zheludev}

\affiliation{Neutron Scattering and Magnetism, Institute for Solid
State Physics, ETH Zurich, Switzerland,  8093 Z\"{u}rich,
Switzerland}

\date{\today}

\begin{abstract}
Fine details of the excitation spectrum of the two-dimensional
spin-gap magnet PHCC are revealed by electron spin resonance
investigations. The values of anisotropy parameters and the
orientations of the anisotropy axes are determined by accurate
measurements of the angular, frequency-field and temperature
dependences of the resonance absorption. The properties of a
spin-gap magnet in the vicinity of critical field are discussed in
terms of sublevel splittings and $g$-factor anisotropy.

\end{abstract}

\keywords{low-dimensional magnets, spin-gap magnets, electron spin
resonance}

\pacs{75.10.Kt, 76.30.-v}

%75.10.Kt    Quantum spin liquids, valence bond phases and related phenomena
%76.30.-v Electron paramagnetic resonance and relaxation

\maketitle

\section{Introduction}
Piperazinium hexachlorodicuprate (C$_4$H$_{12}$N$_2$)(Cu$_2$Cl$_6$),
abbreviated as PHCC, was studied recently as a good test example of
a two-dimensional Heisenberg
antiferromagnet.\cite{broholm-njp2007,broholm-prb64,stone-nature}
Inelastic neutron scattering experiments revealed that each magnetic
ion interacts with at least six in-plane neighbours. The strongest
exchange integral value is approximately 1.3~meV. The complex
geometry of the exchange bonds can be envisioned as a set of
strongly coupled antiferromagnetic ladders running along the
$\vect{a}$ axis (from here on we use the axes notation of
Ref.\onlinecite{structure}) of the triclinic crystal while some of
the weaker exchange bonds are frustrated.\cite{broholm-prb64} The
magnetic ground state of PHCC is a non-magnetic singlet separated
from the excited triplet states by an energy gap of
$\Delta=1.02$~meV. \cite{broholm-njp2007}

Due to the presence of the gap PHCC remains in the disordered
spin-liquid state down to temperatures well below the exchange
integral scale. At low temperatures the magnetic properties of this
system can be considered as those of a gas of weakly-interacting
triplets. The application of an external field lowers the energy of
one of the triplet substates and, at a certain critical field, a
quantum phase transition to a field-induced antiferromagnetic state
occurs. The closing of the gap and the formation of magnetic order
above the critical field of $\approx8$~T was confirmed by magnetic
measurements as well as by calorimetric and neutron scattering
techniques \cite{broholm-njp2007}. Additionally, this magnet was
used as a test probe for the quasiparticle breakdown in a quantum
spin liquid \cite{stone-nature}.

Magnetization measurements revealed an anisotropy of the critical
field \cite{broholm-njp2007}. This anisotropy points to deviations
from the Heisenberg model. Such deviations are important for the
physics of a spin-gap magnet, especially around the quantum phase
transition. Anisotropic interactions are known to cause a gap
reopening in fields above the phase transition and they could also
affect the quasiparticle decay by allowing otherwise forbidden
processes.

In this paper we present results of electron spin resonance (ESR)
investigations of PHCC. Our study reveals the presence of
anisotropic interactions leading to a $g$-factor anisotropy and a
splitting in the triplet sublevels. We have determined magnitudes of
these anisotropic effects and the orientations of the corresponding
axes. Additionally, we address the question of the correct
implementation of a macroscopic theory \cite{Affleck,farmar} to the
case of a spin-gap magnet with a relevant $g$-factor anisotropy.

\section{Experimental details and samples}

The samples were grown from saturated solutions of PHCC. The
saturated solution was prepared by adding of piperazine
C$_4$H$_{10}$N$_2$ (from Sigma Aldrich), dissolved in minimal amount
of concentrated HCl to CuCl$_2 \cdot2$H$_2$O (99.99\% purity from
Sigma Aldrich), dissolved in minimal amount of concentrated HCl, at
a 4:1 molar ratio of CuCl$_2$ to C$_4$H$_{10}$N$_2$.

The main set of samples for the ESR measurements was grown by cooling the
filtered saturated solution in the refrigerator. After several days
numerous as-grown crystals have appeared and were naturally shaped
as flat parallelograms of transparent orange-brown color. The typical
sample size was $3-5$~mm along the longest side of the parallelogram.
The crystallographic symmetry was checked by X-ray diffraction and
found to be triclinic with lattice parameters and angles in agreement with the literature
\cite{structure}. The plane of the as-grown crystals was perpendicular to the $\vect{a}^*$ axis of the reciprocal
lattice. The long and short side of the parallelogram was parallel to the $\vect{c}$
and the $(\vect{b}+\vect{c})$ direction, respectively. This natural shape of the crystals
allowed an easy orientation of the crystals in the magnetic field $\vect{H}$ with $\vect{H}||\vect{a}^*$, $\vect{H}\perp \{\vect{a}^*,\vect{c}\}$
or $\vect{H}\perp \{\vect{a}^*,(\vect{b}+\vect{c})\}$.

Additionally, a larger sample was grown from the seed by a
temperature gradient method. This sample has a well developed plane
normal to the $\vect{a}^*$ axis and has its longest dimension
parallel to $\vect{c}$-axis. For reasons related to the sample space
configuration this sample was used for measurements with
$\vect{H}||\vect{a}^*$ and $\vect{H}\perp \{\vect{a}^*,\vect{c}\}$
only. Samples of deuterated PHCC used for the specific heat
measurements were grown from the seed in a similar way.

The samples were characterized by static magnetization measurements
with a Quantum Design MPMS system. The concentration of impurities
was estimated from the low-temperature susceptibility assuming
$g=2.0$ and $S=1/2$ and was less than $4\times10^{-4}$ paramagnetic
impurities per copper ion.

The technique of electron spin resonance is a powerful tool for the
study of low energy dynamics of a spin-gap magnet. It can access
inter-triplet transitions and measures directly the differences
between triplet sublevels. Characteristic orientational,
frequency-field and temperature dependences of the resonance
absorption spectra provide information on anisotropic interactions,
zero field splittings and the $g$-factor anisotropy.

In case of PHCC additional complications and challenges arise from
the low (triclinic) symmetry of the crystals. Due to this low
symmetry, neither the $g$-tensor nor the anisotropy tensor are fixed
to any of the crystal axes. This means that the axes orientation for
all anisotropic contributions should be considered as arbitrary and
mutually independent. To solve this problem we have taken a series
of absorption spectra at different temperatures and at different
orientations of the applied magnetic field.

The ESR measurements were carried out in the frequency range from
9GHz to 120GHz and at temperatures down to 0.4~K. X-band (9.4~GHz)
measurements above $T=2.5$~K were done with a Bruker EleXsys
spectrometer equipped with a He-flow type cryostat and an automated
sample goniometer. For higher frequency measurements a set of
homemade transmission type ESR spectrometers equipped with $6-8$~T
cryomagnets and microwave oscillators covering the frequency range
up to 120~GHz were used. Measurements below 1~K were performed with
a home made ESR spectrometer equipped with $^3$He-pumping cryostat
and a 12~T cryomagnet.

Low-temperature specific heat measurements in an applied magnetic
field up to 14~T were performed with a Quantum Design PPMS system
equipped with a dilution refrigerator.

\section{Experimental results}

\subsection{Temperature evolution of ESR spectra}

\begin{figure}
\centering
  \epsfig{file=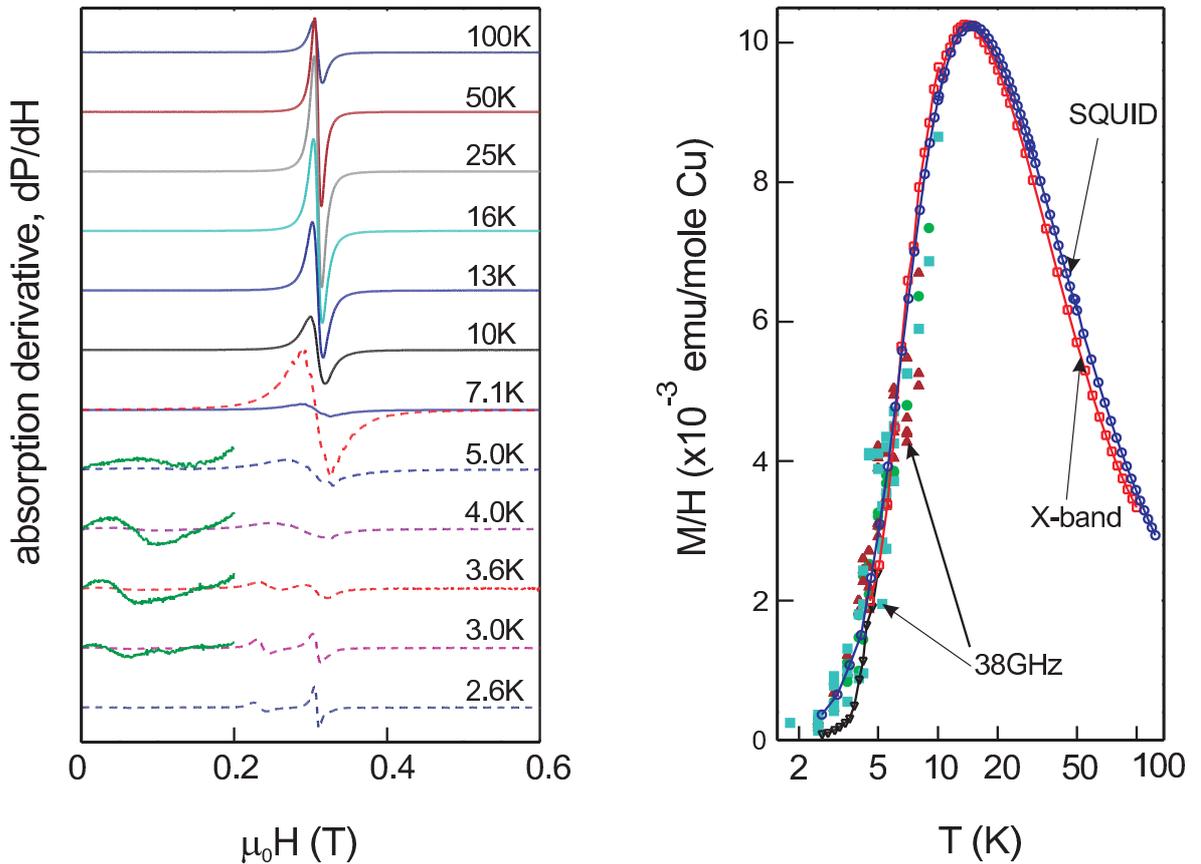, width=\figwidth, clip=}\\
  \caption{(color online) Left panel: Temperature evolution of
  X-band ESR absorption derivative (arb. units, $f=9.4$~GHz, orientation close to $\vect{H}\perp\{
  \vect{a}^*,\vect{c}\}$). Dashed lines are y-magnified by a factor of 10 with
  respect to the raw data. Low-field part of low-temperature
  absorption curves is y-magnified by a factor of 100 with
  respect to the raw data. Right panel: Temperature dependence of the
  static susceptibility (filled circles, denoted as "SQUID") and integrated intensity of ESR
  absorption (filled squares, denoted as "X-band", from the analysis of
  absorption at the frequency of $9.4$~GHz; other symbols, denoted as
  "38~GHz", from the analysis of absorption in different
  orientations at frequencies close to 38~GHz).}\label{fig:x-band&i}
\end{figure}

\begin{figure}
  \centering
  \epsfig{file=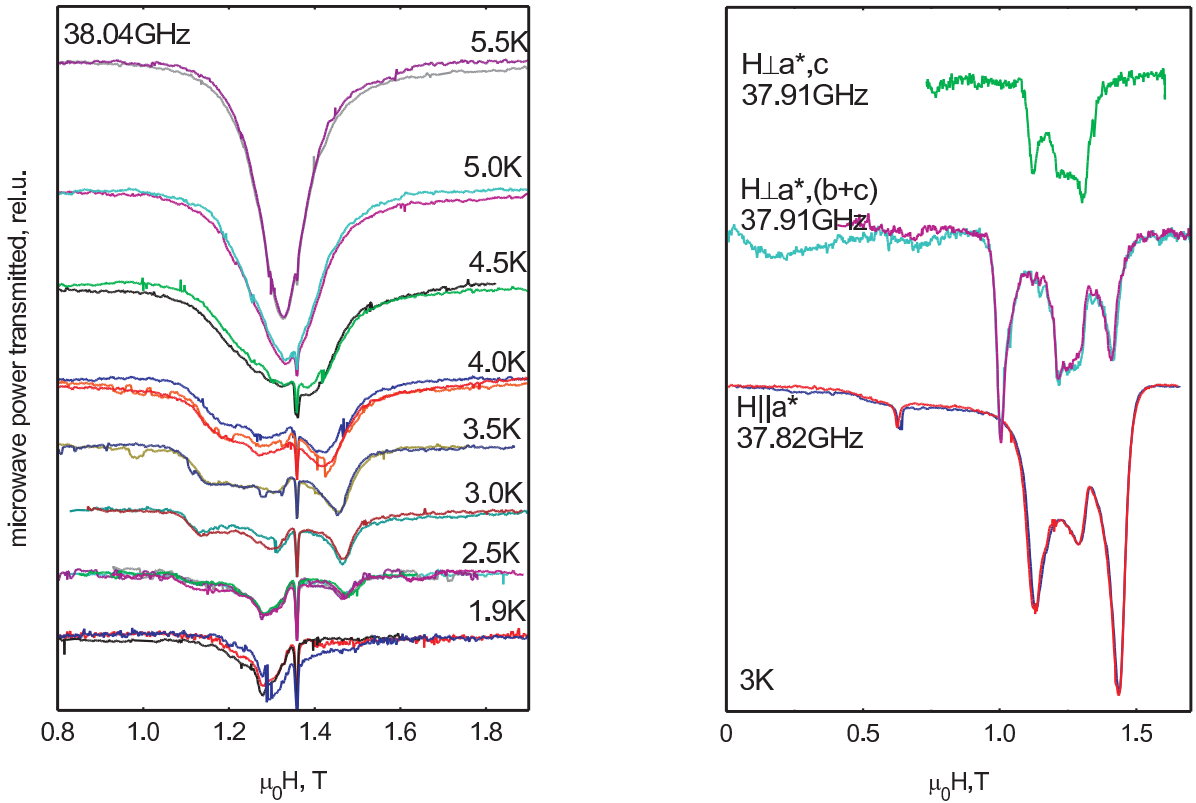, width=\figwidth, clip=}
  \caption{(color online) Left panel: Temperature dependence of ESR absorption
  ($f=38.04$~GHz, $\vect{H}||\vect{a}^*$). Right panel: ESR absorption spectra
  for different orientations at 3.0~K.}\label{fig:scans(t)}
\end{figure}

The temperature evolution of the ESR absorption is illustrated in Figures
 \ref{fig:x-band&i} and {}\ref{fig:scans(t)}. At high temperatures we
observe a single-component ESR absorption with a $g$-factor close to
2, which is typical for Cu$^{2+}$ ions. Below approximately
$10-15$~K the intensity of the resonance quickly decreases. As
expected for a spin-gap magnet this decrease can be scaled with the
decrease of the static susceptibility (see right panel of Figure
\ref{fig:x-band&i}).

As the intensity of the absorption decreases with cooling the
absorption line broadens and, around 5~K, it splits into several
components. This splitting is anisotropic (see right panel of Figure
\ref{fig:scans(t)}). One can distinguish several components that
continue to fade out with cooling and one component, located close
to a $g=2.0$ resonance field, with an increasing intensity upon
cooling. The fading components can be ascribed to a resonance of
triplet excitations, while an irregularly shaped component with
increasing intensity corresponds to defects and impurities.

The magnitude of the splitting is temperature dependent but appears
almost constant for temperatures lower or equal to 3~K (see Figures
\ref{fig:x-band&i} and {}\ref{fig:scans(t)}). Because of the rapid
decrease of intensity of the split spectral features below 3~K we
performed all measurements for the determination of the splitting
parameters at 3~K.

\subsection{Frequency-field diagrams}
\begin{figure}
  \epsfig{file=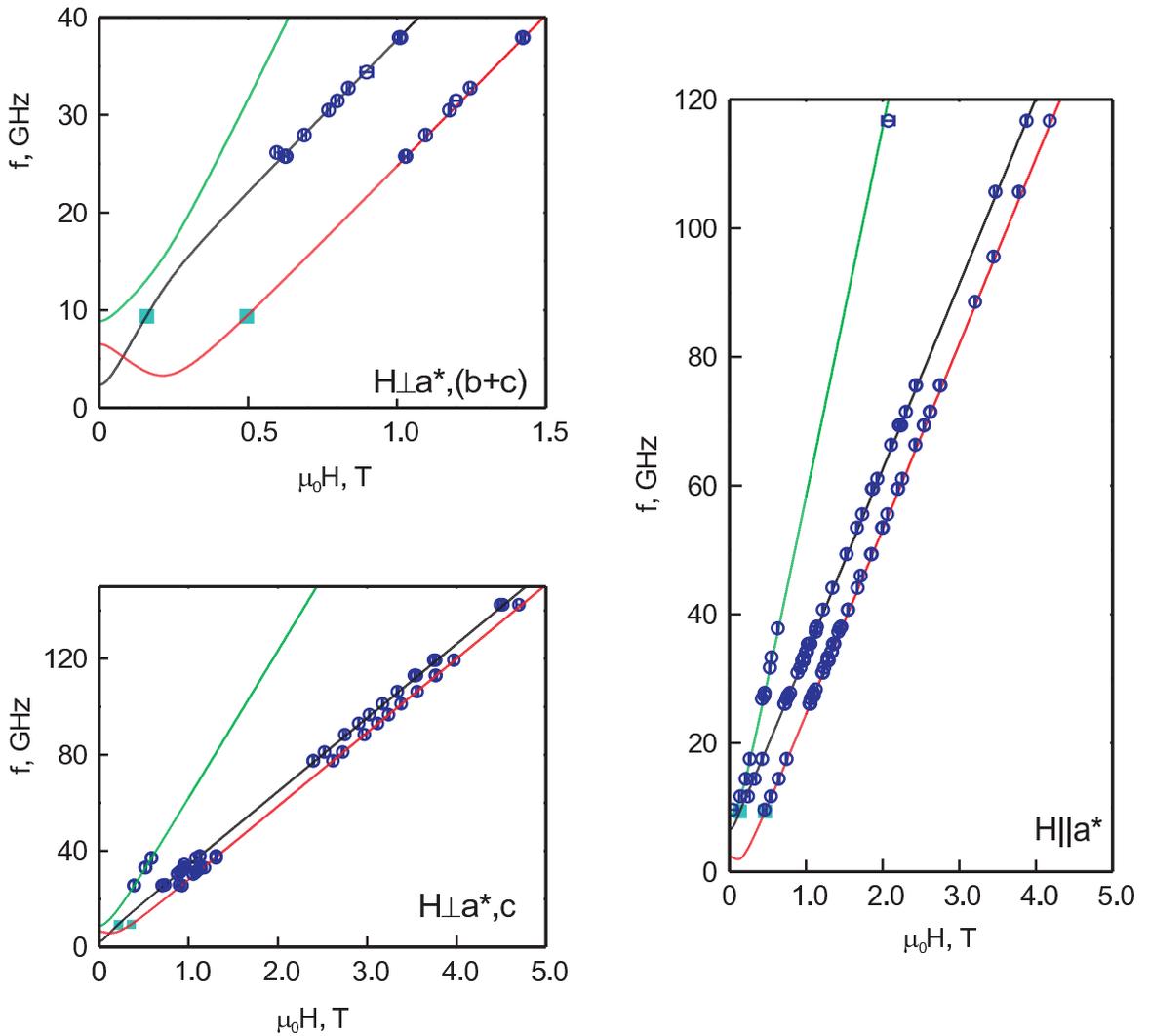, width=\figwidth, clip=}\\
  \caption{(color online) Frequency field diagrams for different orientations of
  the applied field at $T=3.0$~K. Open circles: high frequency
  measurements at fixed orientations. Closed squares: resonance
  field positions for corresponding orientations extracted from
  X-band (9.4GHz) orientational dependences. Curves:
  perturbative model calculations.}\label{fig:f(h)}
\end{figure}

Above the splitting temperature where the ESR absorption consists of
a single component the measurements at different frequencies proved that the resonance field can be described by an effective $g$-factor: for $\vect{H}||\vect{a}^*$ $g=2.05\pm0.01$, for $\vect{H}\perp \{\vect{a}^*, \vect{c}\}$ $g=2.206\pm0.002$ and for $\vect{H}\perp \{\vect{a}^*,
(\vect{b}+\vect{c})\}$ $g=2.236\pm0.002$.

Frequency-field diagrams for the resonance fields measured at 3.0K
are shown on the Figure \ref{fig:f(h)}. For all three chosen
orientations the two main absorption components are located to the
left and to the right of the high-temperature resonance position.
Another, weaker component located at approximately half of the first
resonance field (see also right panel of Figure \ref{fig:scans(t)})
was observed in the measurements with the larger sample for
$\vect{H}||\vect{a}^*$ and $\vect{H}\perp\{ \vect{a}^*,\vect{c}\}$.
The slope of the main components is  determined by  corresponding
$g$-factor value. The splitting of these components is orientation
dependent and is equal to 9.2~GHz, 5.5~GHz and 12.8~GHz for
$\vect{H}||\vect{a}^*$, $\vect{H}\perp \{\vect{a}^*, \vect{c}\}$ and
$\vect{H}\perp \{\vect{a}^*, (\vect{b}+\vect{c})\}$,
correspondingly.

\subsection{Orientational dependences}
\begin{figure}
  \epsfig{file=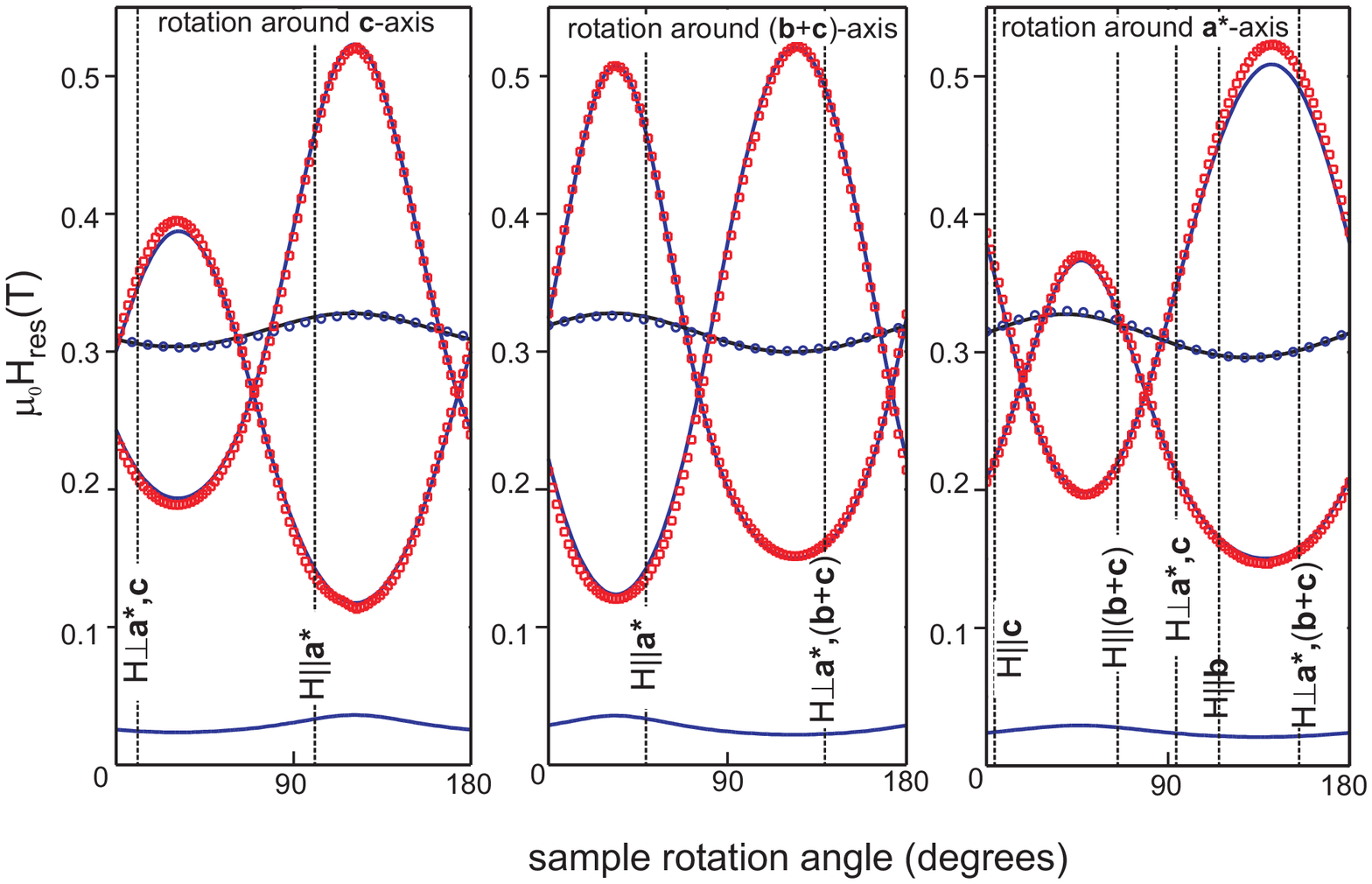, width=\figwidth, clip=}\\
  \caption{(color online) Orientational dependences of resonance field in X-band
  (9.4~GHz) experiments. Selected field directions with respect to
  the crystal axes are shown as determined from the fit. Circles, squares: experimental data at 25~K and 3.0~K, respectively. Lines: best fit by the perturbative model described in the text.}\label{fig:H(phi)}
\end{figure}

The angular dependences of the ESR absorption were accurately
measured at X-band frequency (9.4~GHz) for $T=3.0$~K (below the
splitting temperature) and for $T=25$~K (above the splitting
temperature) where the linewidth shows a minimum. Three
non-orthogonal rotation axes were chosen for different experiments,
namely rotations around the $\vect{a}^*$ reciprocal space axis,
around the $\vect{c}$ axis and around the $(\vect{b}+\vect{c})$
direction. The samples were fixed with paraffine inside suprasil
glass tubes, with the chosen rotation axis parallel to the tube
axis. The accuracy of this sample mounting technique was limited
with respect to the rotations around the tube axis and, thus, the
unavoidable angle shift was later taken as a fit parameter.

The positions of the resonance fields shown in Figure
\ref{fig:H(phi)} were determined from fitting the spectra by a
single or a multi-component absorption, assuming a Lorentzian shape
for each of the components. For the data at 3.0~K we did not include
the weak low-field component (see the line below 0.2~T in the left
panel of Fig. \ref{fig:x-band&i}), since this line is broad and
distorted in shape, which makes a precise determination of the
resonance field unreliable.

\subsection{Low temperature ESR measurements}
To check for possible singlet-triplet transitions and
antiferromagnetic resonance in the field-induced ordered phase we
performed additional measurements at the temperatures down to 450~mK
using the large sample with $\vect{H}\|\vect{a}^*$. At frequencies
of 27~GHz and 33~GHz and for fields up to 12T we did not observe any
other resonance absorption at 450~mK besides that due to defects. On
heating above 1.5~K a split ESR absorption, similar to that
described in the above subsections, was observed.

\subsection{Critical field measurements}

\begin{figure}
  \epsfig{file=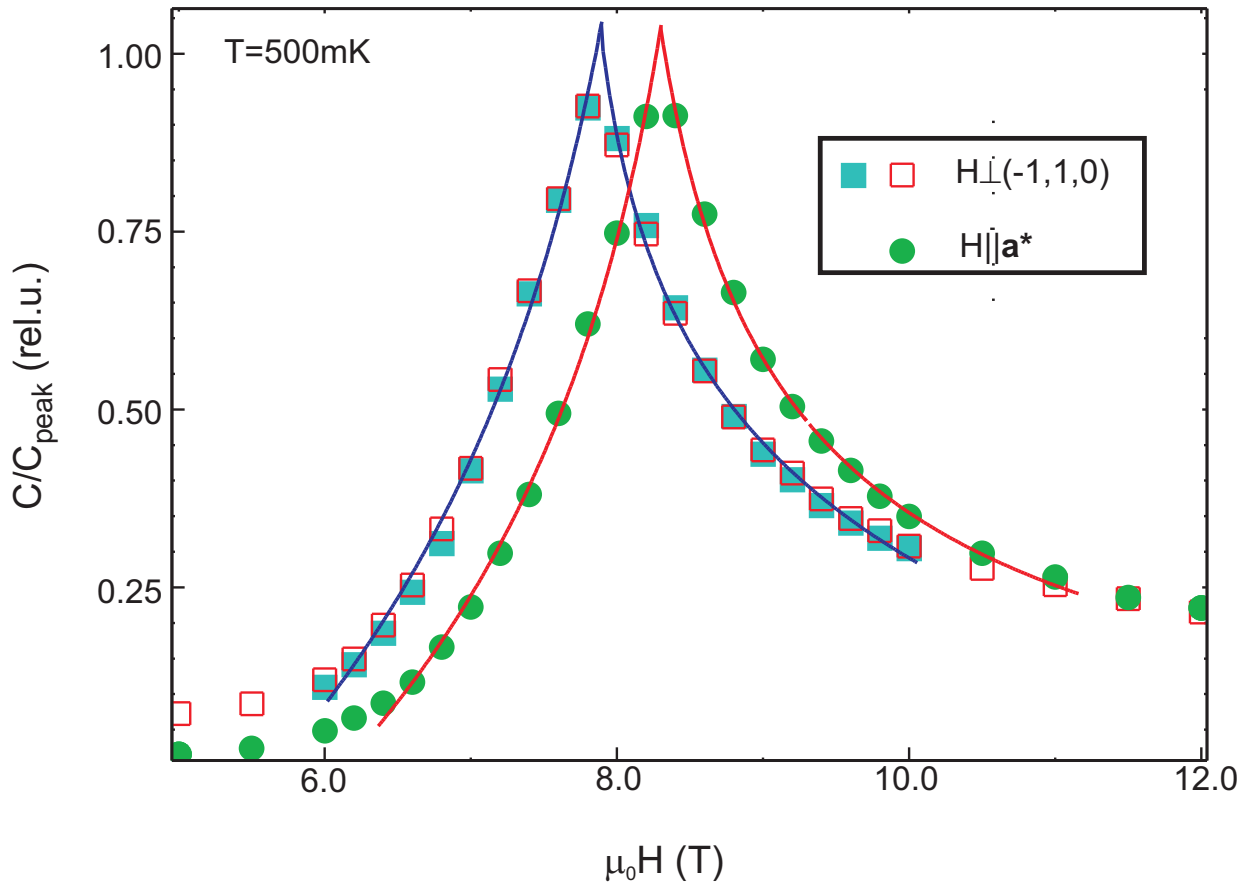, width=\figwidth, clip=}\\
  \caption{(color online) Field dependences of the specific heat of
  deuterated PHCC at different orientations and at $T=0.5$~K. Curves are guides to the eye.  }\label{fig:specheat}
\end{figure}
In order to determine the critical fields at different orientations
the field dependence of the specific heat was measured at 500~mK
(Figure \ref{fig:specheat}). These measurements were done on
deuterated samples of PHCC while non-deuterated samples were used
for ESR experiments. A sharp maximum of the specific heat marks the
transition to the field-induced ordered phase. The determined values
of the critical fields are: for $\vect{H}||\vect{a}^*$
$\mu_0H_c=8.29\pm0.05$T, for $\vect{H}\perp(-1,1,0)$
$\mu_0H_c=7.90\pm0.07$T.

\section{Discussion}
This section is built as follows: first, we will qualitatively
describe our data and will show that our observations reflect the
effects of anisotropic interactions; second, we will make a
quantitative analysis of the ESR data and will estimate the strength
of these anisotropic interactions; finally, we will discuss the
relations between the found anisotropic properties at low field with
the known anisotropic properties at the critical field.

\subsection{Qualitative description}

The observed evolution of the ESR absorption spectra is typical for
a spin-gap magnet with triplet levels split by an effective crystal
field. A similar behaviour was observed earlier for other spin-gap
magnets, e.g. TlCuCl$_3$ \cite{glazkov-tlcucl3} and NTENP
\cite{glazkov-ntenp}.

The decrease of the absorption intensity with temperature
corresponds to the decrease of the population numbers of gapped
excitations. At high temperature the population numbers are high and
excited quasiparticles interact with each other switching on the
exchange narrowing mechanism. This causes a single resonance line
observed at high temperature. At low temperatures excitations are
rare and can be considered as a gas of noninteracting
quasiparticles. Hence, their spectrum can be affected by anisotropic
interactions, which can be described as an effective crystal field
acting on $S=1$ quasiparticles. These interactions lift the
degeneracy of the triplet sublevels at $H=0$, causing splitting of
the thermoactivated ESR absorption spectrum into several components.
This effect is similar to the well known splitting of energy levels
of an $S=1$ ion in a crystal \cite{Abragam}. The line broadening
marks a crossover from the high-temperature, exchange-narrowed limit
to the low-temperature single-particle limit.

Besides the thermoactivated ESR absorption which is caused by the
resonance transitions between triplet sublevels an observation of
singlet-triplet transitions or an antiferromagnetic resonance above
the critical field could be possible. Both of these transitions were
observed in the above-mentioned spin-gap systems TlCuCl$_3$ and
NTENP. However, the intensity of the dipolar singlet-triplet
transition should be zero in the exchange approximation. Thus the
observation of this transition should require in particular a
symmetry-breaking of the ground state by anisotropic interactions.
The absence of the singlet-triplet transition in our experiment
shows that the coupling of these states is small. The absence of the
resonance absorption in the field induced ordered phase even up to
12T at 450mK (at $H=10$T $T_N\approx 2.5K$ \cite{broholm-njp2007})
provides 30GHz (or 0.15meV) as an upper estimate of the gap.

\subsection{Application of a perturbative model for the description of ESR data}
\begin{figure}
  \epsfig{file=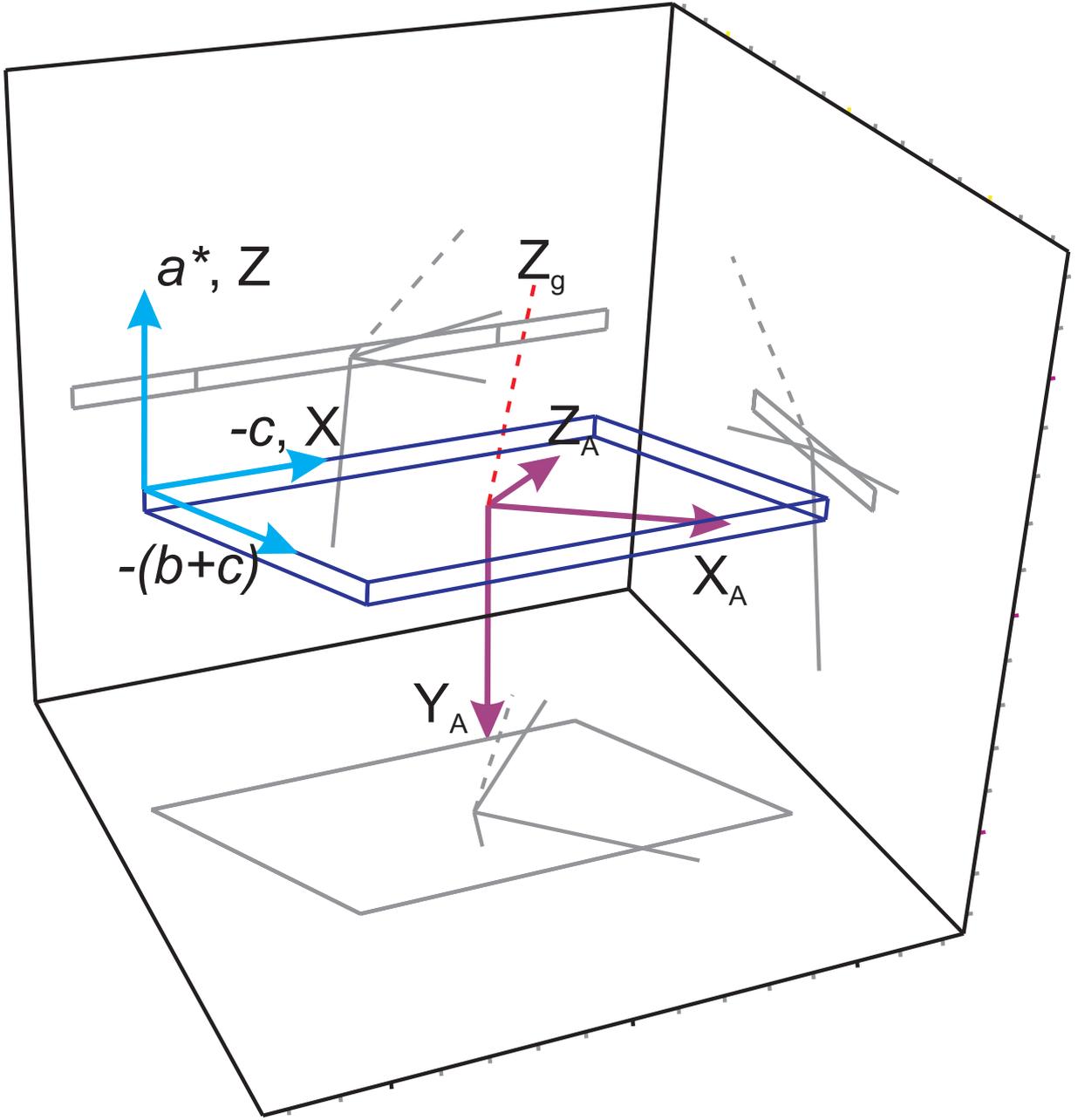, width=\figwidth, clip=}\\
  \caption{(color online) Orientation of the cartesian basis used for calculations
  with respect to the crystal axes and orientations of anisotropy
  and $g$-tensor axes found from the modeling.}\label{fig:model axes}
\end{figure}

\begin{figure}
  \centering
  \epsfig{file=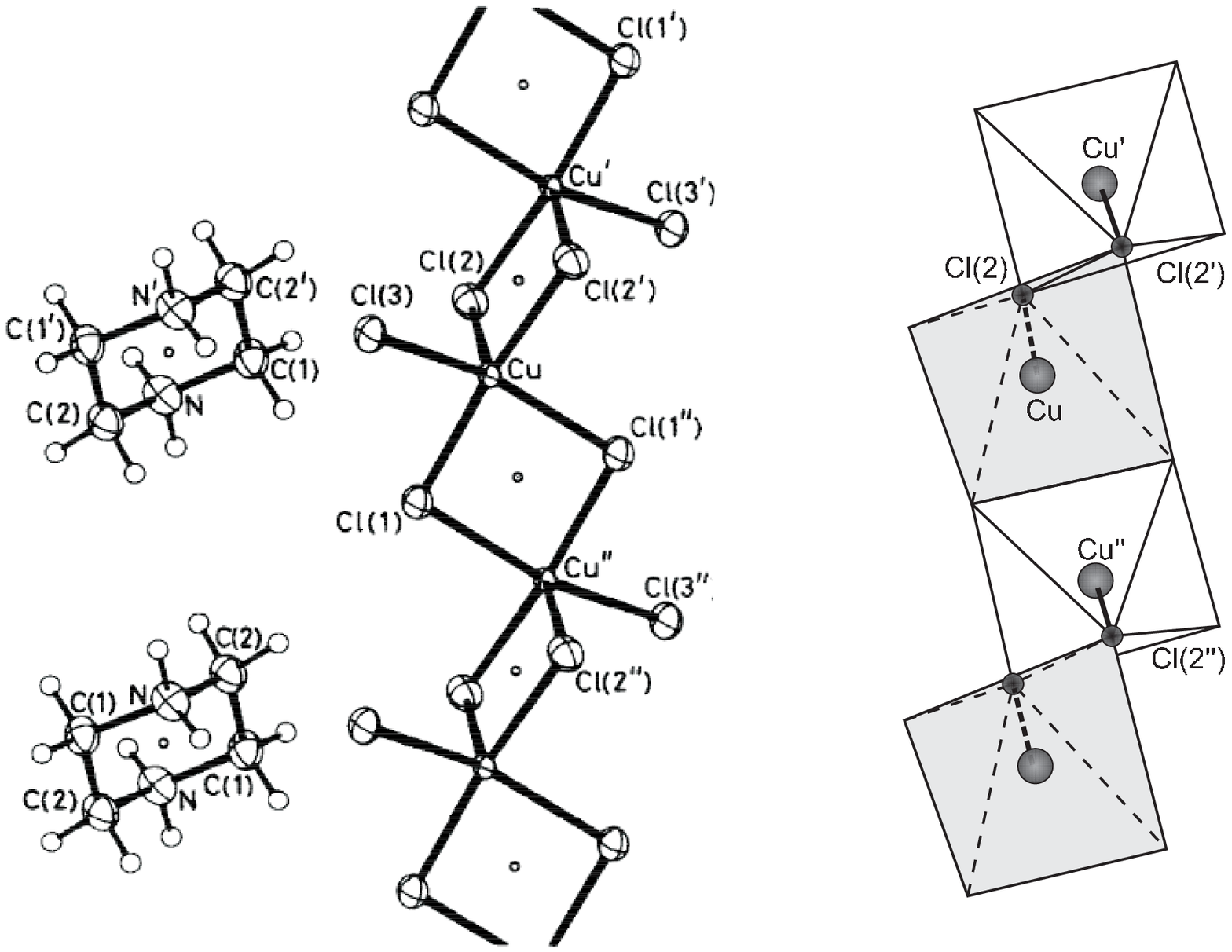, width=\figwidth, clip=}
  \caption{Left: Fragment of the PHCC crystallographic structure (from
  Ref.\onlinecite{structure}). Vertical direction corresponds to
  the crystallographic $\vect{c}$ axis. Right: Orientations of the
  distorted chlorine pyramids. Positions of copper and apical
  chlorine ions are shown. Visible bases of the pyramids are
  shadowed in grey. }\label{fig:struct}
\end{figure}

Different models are used for the description of the low energy
dynamics of a spin gap magnet in magnetic field. A fermionic
field-theory model was proposed for 1D systems by Tsvelik
\cite{tsvelik}. Alternatively, a bosonic field-theory model was
proposed by Affleck \cite{Affleck} and later independently developed
as a macroscopic model by Farutin \cite{farmar}. Particular systems
can be treated by microscopic approaches \cite{Kolezhuk}. However,
only close to the critical field the differences of all these models
become important. Since we can not observe singlet-triplet
transitions on PHCC, we can not look for these fine differences and
can not reliably judge on the applicability of different models from
the ESR data alone. On the other hand, since our data correspond to
the low-field limit (the critical field is around 8~T, while all
X-band resonance modes are below 1~T and most of the higher frquency
data are below 4~T) we can apply a perturbative approach for the
description of the observed ESR data. It will yield the values of
zero-field splitting of the energy levels and the orientations of
all relevant anisotropy axes which can be used, if necessary, to
obtain parameters of other models.

This approach considers triplet excitations for a non-interacting
$S=1$ particle moving in a stationary effective crystal field. The
effective spin Hamiltonian for this particle is:

\begin{equation}\label{eqn:ham}
    \ham=\Delta+\vect{H}\cdot\widetilde{g}\cdot\svect+D\sz^2+E(\sx^2-\sy^2)
\end{equation}

\noindent where $\Delta$ is an energy gap, $\widetilde{g}$ is a
$g$-tensor, $D$ and $E$ are the effective anisotropy parameters
and $\vect{X}_A$, $\vect{Y}_A$, $\vect{Z}_A$ are the anisotropy
axes. Due to the triclinic crystal symmetry the orientation of
the anisotropy axes $\vect{X}_A$, $\vect{Y}_A$, $\vect{Z}_A$ is
arbitrary with respect to the crystal. Moreover, the $g$-tensor axes
$\vect{X}_g$, $\vect{Y}_g$, $\vect{Z}_g$ are also arbitrary with
respect both to the crystal and to the anisotropy axes. We neglect
here the $\vect{k}$-dependence of the Hamiltonian parameters $D$ and
$E$, since at low temperatures only the bottom of the spectrum is
populated, while at high temperatures the exchange narrowing cancels
out the effects of these terms.

The perturbative model can be solved for eigenenergies for any given
magnetic field. The differences of the eigenenergies will correspond to the
resonance frequencies which can then be compared with the
experiment. We used a standard least-squares routine to minimize
the model deviation from the experimental data.

Angular dependences of the X-band resonance fields at 3~K and 25~K
were used for the determination of the Hamiltonian parameters.
Additionally, the values of the $g$-factor and the low-temperature
splittings measured in the high-frequency experiments with
well-defined sample orientation were used as anchors to correct for
the angle shifts due to the sample mounting. The basis $\vect{X}$,
$\vect{Y}$, $\vect{Z}$ with $\vect{Z}||\vect{a}^*$ and
$\vect{X}||(-\vect{c})$ was selected to describe the orientation of
the $g$-tensor and the anisotropy axes with respect to the crystal
(Figure \ref{fig:model axes}).

This problem has 14 parameters: the anisotropy constants $D$ and
$E$, the Euler angles of the anisotropy axes $\alpha_A$, $\beta_A$
and $\gamma_A$, three main values of the $g$-tensor, the Euler
angles of the $g$-tensor axes, and, finally, three uncontrolled
shift angles $\epsilon_{1,2,3}$ in the X-band rotation experiments.
It turned out, that this general fit yields two of the $g$-tensor
components very close to each other (best fit $g$-factor values
2.272, 2.062 and 2.039). Consequently, the fit became unstable and
could not well distinguish the orientations of the $g$-tensor
differing by rotation around high-$g$-value axis. Thus, we reduced
the number of model parameters to 12 by assuming an axial $g$-tensor
that can be described by two main values and by two polar angles of
the $g$-tensor axis. The stability of the best fit solution was
confirmed by running numerous fitting procedures with randomly
varied initial approximations of the model parameters.

\begin{table}
\caption{Best fit parameters in the perturbative
model.\label{tab:bestfit}}
\begin{tabular}{|c|c|c|}
  \hline
  anisotropy&D, MHz & $-7900\pm280$ \\
  constants& E, MHz & $-1340\pm190$ \\
  \hline
  $g$-tensor&$g_{||}$ & $2.280\pm0.015$ \\
  main values&$g_\perp$ & $2.048\pm0.007$ \\
  \hline
  Euler angles&$\alpha_A$ & $(-43.5\pm1.8)^\circ$ \\
  of anisotropy &$\beta_A$ & $(-105.6\pm1.5)^\circ$ \\
  axes& $\gamma_A$ & $(-12.5\pm7)^\circ$ \\
  \hline
  Polar angles of&  $\Theta_g$ & $(105\pm3)^\circ$ \\
  $g$-tensor axes&$\phi_g$ & $(54.9\pm1.5)^\circ$ \\
  \hline
  shift&$\epsilon_1$ & $(-10.9\pm3.5)^\circ$ \\
  angles&$\epsilon_2$ & $(41\pm2.5)^\circ$ \\
  &$\epsilon_3$ & $(-4.0\pm3.5)^\circ$ \\
  \hline
\end{tabular}
\end{table}

The best fit parameters are shown in the Table \ref{tab:bestfit} and
the corresponding frequency-field and angular dependences are shown
in the Figures \ref{fig:f(h)}, \ref{fig:H(phi)}. The orientation of
the anisotropy axes and $g$-tensor axis are illustrated at the
Figure \ref{fig:model axes}. The sign of anisotropy constants can be
unambiguously determined by comparing low-temperature intensities of
the split absorption components in different orientations (Figure
\ref{fig:scans(t)}): a more intense absorption corresponds to the
transition from the lower sublevel.

The found axiality of the $g$-tensor can be explained by the details of
the microscopic structure of PHCC.\cite{structure} Each copper ion is
surrounded by five chlorine ions, forming a distorted square-based
pyramid (Figure \ref{fig:struct}). The main axis of this pyramid is the
same for all copper ions (differently oriented pyramids are
connected by inversion). The direction of the vector linking the copper
to the apical chlorine (position Cl(2) in the denominations of
Ref.\onlinecite{structure}) is very close to the found direction of
the $g$-tensor axis: its polar angles in the chosen reference frame
are $\Theta_{Cu-Cl}=111^\circ$ and $\phi_{Cu-Cl}=57^\circ$.

\subsection{Anisotropic corrections and critical field.}
\begin{table}
\caption{Comparison of the measured parameters of the excitation
spectrum and calculated parameters in different models. Values of
$g$-factor are given for reference as calculated for particular
orientations using a best fit perturbative model. For the precision of
the model estimations see footnote \footnote{Estimated errors of the model
calculations: Absolute value of zero field energy $\pm 2$GHz.
Difference of sublevels energies $\pm 0.4$GHz. Critical fields: $\pm
0.08$T. This error is mostly due to the $g$-factor uncertainty.}.
}\label{tab:compare}
\begin{tabular}{|c|c|c|c|c|}
    \hline
    &Experiment&$g$-factor&pertur-&macro-\\
    &   & anisotropy& bative  & scopic\\
    &   & only& model & model\\
    \hline
    Zero-field &  &   &233.6   & 234.0  \\
    energies,&247\footnote{Ref.\onlinecite{broholm-njp2007}, deuterated PHCC, inelastic neutron scattering, T=60mK}  & 235.6  &236.2  &236.6 \\
    GHz   &   &   &242.8   & 243.2   \\
    \hline
    \hline
    $\mu_0 H_c$,T:\hfill& & & & \\
    \hline

    $\vect{H}\perp(-1,1,0)$&$7.90\pm0.07$\footnotemark[3]&7.71 & 7.74& 7.70\\
    ($g=2.183$)& & & & \\
    \hline

    $\vect{H}||\vect{a}^*$&$8.29\pm0.05$\footnote{present work, deuterated PHCC, C(H) at T=0.5K} & & & \\
    ($g=2.064$)&$8.03\pm0.07$\footnote{Ref.\onlinecite{broholm-njp2007}, magnetization measurements at 0.46K} &8.15 & 8.27 &8.19 \\
    \hline
    $\vect{H}||\vect{b}$&$7.47\pm0.01$\footnote{Ref.\onlinecite{broholm-njp2007}, deuterated PHCC, elastic neutron scattering, T=65mK} &7.47 &7.47 & 7.47   \\
    ($g=2.253$)& $7.58\pm0.03$ \footnotemark[2]  &   &   &   \\
    & $7.40\pm0.08$ \footnotemark[4]  &   &   &   \\
    \hline
    $\vect{H}||\vect{c}$&$7.6\pm0.2$\footnotemark[4] &7.93 & 7.96 &7.89 \\
    ($g=2.122$)& & & & \\
    \hline

\end{tabular}

\end{table}

Our specific heat measurement and earlier measurements of other
authors demonstrate the anisotropy of the critical field of about
1~T (see Table \ref{tab:compare}). Additionally, neutron scattering
experiments \cite{broholm-njp2007,stone-prl96} have determined the
orientation of the order parameter above the critical field for
$\vect{H}||\vect{b}$: the order parameter was found to lie within
$(7\pm4)^\circ$ of the $(ac)$ plane with an in-plane component
within $(6\pm4)^\circ$ of the $a$-axis. Both of these effects are
determined by the anisotropic interactions in the spin system and
thus provide an independent check for the anisotropy parameters
found above.

The critical field depends strongly on temperature: its values at
65mK and at 2.8K differ by $\approx 2$T \cite{broholm-njp2007}.
Thus, it is necessary to ascertain whether the anisotropy parameters
determined at 3.0K can be used in the analysis of the low
temperature data or not.

The temperature dependence of the critical field originates from the
renormalization of the energy spectrum due to the repulsive
interaction of triplet excitations. This renormalization depends on
the triplet population numbers and therefore affects strongly the
energy spectrum in the vicinity of the critical field, where the
energy gap is small. Our ESR measurements, on the contrary, were
performed in small fields where the energy gap value is still high
and the triplet population numbers are small even at 3.0K. This is
confirmed by the experimental observation that the observed
splitting of the resonance absorption does not change significantly
at further cooling below 3K. Therefore, the effect of triplet
repulsion does not affect the values of anisotropy parameters found
in the previous subsection and these values can be used for the
analysis of the low temperature data for the critical field
anisotropy and order parameter orientation.

In the model calculations of this subsection we will use $g$-factor
values and (when necessary) zero-field splittings as determined in
our ESR experiments. Since the gap value $\Delta$ can not be
determined with the similar precision, we will tune it to obtain
correct value of the critical field in the chosen orientation. We
choose the value of the field at which the antiferromagnetic Bragg
reflection appears \cite{broholm-njp2007} at $\vect{H}||\vect{b}$
($\mu_0H_c=7.47\pm0.01$T) as the most reliable anchor for this
purpose.

First, there is a simplest model, suggested in Ref.
\onlinecite{broholm-njp2007}, that treats the  anisotropy of the
critical field as originating from the $g$-factor anisotropy only.
This model was backed by the observation that the  ratios of the
critical field to the saturation field were found to be the same in
different orientations. The values of the $g$-factor in the
particular orientations of the magnetic field calculated from our
model are given in Table \ref{tab:compare}.  One can see that the
highest critical field is indeed observed in the direction with the
smallest $g$-value ($\vect{H}||\vect{a}^*$). The critical field in
the arbitrary orientation can be expressed via the $g$-factor value
as $H_c=(g(\vect{H}||\vect{b})H_c(\vect{H}||\vect{b}))/g$. The
calculated values are in reasonable agreement with the experiment
(see Table \ref{tab:compare}).

However, this simple model is not exact, since we observe zero-field
splitting of energy levels. The effect of zero-field splitting on
the anisotropy of the critical field  can be roughly estimated from
the value of the main anisotropy constant as $\mu_0\Delta H_{A}\sim
hD/(g\mu_{B})\approx0.3$T. This has to be compared with the effects
of the $g$-tensor anisotropy: $\mu_0\Delta H_g\sim(\Delta g/g) \mu_0
H_c\approx 0.9$T. These values are similar and, especially since the
axes of anisotropy and the axes of $g$-tensor do not coincide, it is
not possible to neglect {\em a priori} one of the effects  for an
arbitrary field direction.

As a second test model, we extrapolate the results of the
perturbative model up to the critical field. This extrapolation can
be justified only for one-dimensional systems\cite{zaliznyak}.
However, this model can be considered as a simple approximation
taking into account both the effects of $g$-tensor anisotropy and
zero-field splitting of triplet sublevels. The results of this
extrapolation are also collected in Table \ref{tab:compare}.

Finally, the problem of the triplet level field dependence can be
treated more strictly within the frameworks of a macroscopic
(bosonic) approach\cite{farmar,Affleck}. This model treats the
excitations of a spin-gap magnet as oscillations of the vector field
$\boldeta$ with the Lagrangian

\begin{equation}\label{eqn:lagr}
{\cal{L}}=\frac{1}{2} \left(\dot{\boldeta}+\gamma[\boldeta\times
\vect{H}]\right)^2-\frac{1}{2}A\boldeta^2+{\cal{L}}_{rel}
\end{equation}

\noindent where $A$ is the exchange constant describing the energy
gap, $\gamma$ is the gyromagnetic ratio and ${\cal{L}}_{rel}$ is  a
term containing anisotropic relativistic corrections. The leading
corrections terms at low fields are the effective anisotropy terms,
responsible for the zero field splitting, which can be written in
the reference frame of anisotropy axes $(X_A ,Y_A ,Z_A)$:

\begin{equation}\label{eqn:lagr-rel1}
  {\cal{L}}_{rel1}=\frac{1}{2}b_1
  (\eta_{X_A}^2+\eta_{Y_A}^2-2\eta_{Z_A}^2)+\frac{1}{2}b_2(\eta_{X_A}^2-\eta_{Y_A}^2)
\end{equation}

\noindent where $b_{1,2}$ are the effective anisotropy constants.
These terms allow to describe the zero field splitting, see e.g.
Ref.\onlinecite{glazkov-ntenp}.

To describe additional field effects we have to include here next
order terms. The low symmetry of the crystal allows numerous
relativistic and exchange-relativistic invariants such as
 $\dot{\eta}_\alpha \dot{\eta}_{\beta}$,
$[\boldeta\times\dot{\boldeta}]_\alpha H_\beta$,
$(\vect{H}\boldeta)H_\alpha \eta_\beta$, $H^2\eta_\alpha
\eta_\beta$, $\boldeta^2 H_\alpha H_\beta$ etc. I.e., in general we
have to consider a veritable zoo of possible terms which overburdens
the problem beyond any hope to produce a compact solution.

To simplify this problem we will assume that the main field effects
arise from a single-ion $g$-tensor anisotropy. Thus, due to an
almost tetragonal local symmetry, these effects can be considered as
having an axial symmetry in the $g$-tensor reference frame.
Additionally, we will require that for the model test case of
$b_1=b_2=0$ these terms should provide linear E(H) dependences for
all three modes, which is natural for the microscopic model with
anisotropic single-ion $g$-tensor only.

This leads to the following invariants combination:

\begin{equation}\label{eqn:lagr-rel2}
{\cal{L}}_{rel2}=\xi(\gamma^2 H_{Z_g}^2\boldeta^2-\gamma^2
(\vect{H}\boldeta)H_{Z_g}\eta_{Z_g}+\gamma
H_{Z_g}[\dot{\boldeta}\times\boldeta]_{Z_g})
\end{equation}

\noindent where the parameter $\xi=(g_{||}/g_{\perp}-1)$ reflects
the $g$-factor anisotropy. Up to higher order corrections, equation
(\ref{eqn:lagr-rel2}) is equal to the replacement of the problem
with an anisotropic $g$-tensor $\tilde{g}$ and a certain real field
direction $\vect{H}$ by the formally equivalent problem with the
isotropic $g$-factor $g_0$ and a transformed "effective" field
$\vect{H}_{eff}=(1/g_0)(\tilde{g}\vect{H})$. Three terms in
Eqn.(\ref{eqn:lagr-rel2}) are responsible for different effects and
none can be omitted.

The dynamics equations can be obtained from the final Lagrangian
${\cal{L}}+{\cal{L}}_{rel1}+{\cal{L}}_{rel2}$ using a standard
variational technique with special attention to the frames of
reference (see Appendix). The solution of these equations yields the
field dependences of the energy levels, whose differences would
yield transition frequencies. The parameters of the macroscopic
model can be directly expressed via perturbative model parameters
with only one adjustable parameter $\Delta$:

\begin{eqnarray}
A&=&4\pi^2 (\Delta^2+(\Delta+D+E)^2+(\Delta+D-E)^2)/3\nonumber\\
b_1&=&4\pi^2
(2\Delta^2-(\Delta+D+E)^2-(\Delta+D-E)^2)/6\nonumber\\
b_2&=&4\pi^2((\Delta+D+E)^2-(\Delta+D-E)^2)/2\nonumber\\
\gamma&=&2\pi(g_\perp/2.0)\gamma_0\nonumber\\
\xi&=&(g_{||}/g_\perp-1)\nonumber
\end{eqnarray}
\noindent here $\gamma_0=2.80$GHz/kOe is a free electron
gyromagnetic ratio. The factors of $4\pi^2$ or $2\pi$ appears here
since the dynamics equations (\ref{eqn:dyneqn}) are written for
angular frequency $\omega$, while the parameters of the perturbative
model were expressed in the ordinary frequency units.

The parameters values corresponding to
$\mu_0H_c(\vect{H}||\vect{b})=7.47$T are:
$A=2.236\cdot10^{24}$~1/sec$^2$, $b_1=4.972\cdot10^{22}$~1/sec$^2$,
$b_2=-2.490\cdot10^{22}$~1/sec$^2$, $\gamma/(2\pi)=2.867$GHz/kOe,
$\xi=0.1133$.

We have found that the model curves fits our data perfectly and are
undistinguishable from the perturbative curves at low fields. The
predicted values of critical fields are also gathered in the Table
\ref{tab:compare}.

The macroscopic model allows a straightforward analysis of the order
parameter orientation above the critical field. We have to add to
the final Lagrangian fourth order exchange term $-B\boldeta^4$ which
will determine the magnitude of the order parameter at
$H>H_c$.\cite{farmar} However, as long as all anisotropic terms are
of the same order in $\boldeta$, the orientation of the order
parameter does not depend on its amplitude and, consequently, does
not depend on the value of the exchange constant $B$. Thus, for the
sake of determining the order parameter orientation only, the $B$
constant can be simply thought of as a mathematical meaning to
exclude the divergence of the order parameter above $H_c$. Using the
model parameters described above we have found that the polar angles
of order parameter at $\vect{H}||\vect{b}$ are
$\Theta_{O.P.}=34^\circ$ and $\phi_{O.P.}=-20^\circ$. This
corresponds to an angle of 16$^\circ$ with the $(ac)$-plane and to
an angle of 28$^\circ$ to the $a$-axis. The in-plane component of
the calculated order parameter forms an angle of $23^\circ$ with the
$a$-axis. Although there is a qualitative agreement, the calculated
values differ from those found in the experiment
\cite{broholm-njp2007,stone-prl96}.

All of the models provide a reasonable qualitative agreement with
the experiment. The macroscopic model seems to be physically the
most justified, since it describes correctly the zero-field
structure of energy levels and can be applied up to the critical
field and above. Quantitative disagreements with the experiment can
be caused by some neglected factors: the slight change of anisotropy
constants at cooling below 3K or the effects of additional
anisotropic contributions above $H_c$.

\section{Conclusions.}

A detailed investigation of the low-energy spin dynamics in the
spin-gap magnet PHCC revealed the presence of anisotropic
interactions causing a splitting of the triplet sublevels and
leading to a $g$-tensor anisotropy. The effects of these
interactions were described quantitatively within a perturbative
approach in small fields. The parameters of anisotropy comprising
the main $g$-tensor values, the zero-field splitting of sublevels
and the orientations of corresponding axes with respect to the
crystal were found. Additionally, an application of the macroscopic
(bosonic) model to the case of a spin gap magnet with a relevant
$g$-factor anisotropy was developed.

\acknowledgements

Part of this work was supported by the Swiss National Science
Foundation, under Project 6 of MANEP. One of the authors (V.G.)
acknowledge support from Russian Foundation for Basic Research
(RFBR) Grant No. 09-02-00736-a and from Russian Presidential Grant
for the Leading Scientific Schools No 65248.2010.2.

Authors acknowledge usage of the "Measurement Commander" software by
Th.Kurz (University of Augsburg, Experimental Physiscs V (EKM)) for
the fit of the X-band ESR data.

\appendix*
\section{Dynamics equation for macroscopic (bosonic) model with
axial $g$-tensor below $H_c$.}

The complete Lagrangian of the problem can be obtained from
Eqns.(\ref{eqn:lagr}),(\ref{eqn:lagr-rel1}),(\ref{eqn:lagr-rel2}):

\begin{eqnarray}
 \cal{L}&=&\frac{1}{2}(\dot{\boldeta}+\gamma[\boldeta\times
 \vect{H}])^2-\frac{1}{2}A\boldeta^2+\nonumber \\
 &+&\frac{1}{2}b_1
  (\eta_{X_A}^2+\eta_{Y_A}^2-2\eta_{Z_A}^2)+\frac{1}{2}b_2(\eta_{X_A}^2-\eta_{Y_A}^2)+\nonumber\\
  &+&\xi\gamma^2 H_{Z_g}^2\boldeta^2-\xi\gamma^2
(\vect{H}\boldeta)H_{Z_g}\eta_{Z_g}+\nonumber\\
 &+&\xi\gamma
H_{Z_g} [\dot{\boldeta}\times\boldeta]_{Z_g}
\end{eqnarray}

The same coefficient $\xi$ in the last three terms is a consequence
of the microscopic model that assumes a single-ion $g$-tensor
anisotropy with an axial $g$-tensor. Due to the low symmetry of the
crystal, terms responsible for the effective anisotropy ($b_1$ and
$b_2$ terms) and for the $g$-factor anisotropy ($\xi$-terms) are
written in different cartesian bases. When varying the Lagrangian
over $\boldeta$ these terms have to be recalculated to the same
basis. We used the anisotropy axes basis for the derivation of the
dynamics equation. The resulting dynamics equation on $\boldeta$ is:

\begin{eqnarray}
&&\ddot{\boldeta} +2
\gamma[\dot{\boldeta}\times\vect{H}]-\gamma^2H^2\boldeta+
\gamma^2\vect{H}(\boldeta\cdot\vect{H})+A\boldeta- \nonumber
\\
&-& b_1 \left(
    \begin{array}{c}
    \eta_{X_A}\\ \eta_{Y_A}\\ -2\eta_{Z_A}\\
    \end{array}
    \right)
-b_2 \left(
    \begin{array}{c}
    \eta_{X_A}\\ -\eta_{Y_A}\\ 0\\
    \end{array}
    \right)
-\xi\gamma^2(\vect{H}\vect{Z}_g)^2\boldeta+\nonumber\\
&+&\xi\gamma^2(\vect{H}\vect{Z}_g)((\boldeta\vect{Z}_g)\vect{H}+(\boldeta\vect{H})\vect{Z}_g)-\nonumber\\
&-&2\xi\gamma(\vect{H}\vect{Z}_g)[\vect{Z}_g\times\boldeta]=0.
    \label{eqn:dyneqn}
\end{eqnarray}

This equation yields a secular equation for the precession circular
frequencies $\omega_{1,2,3}$ that describes the field dependence of
the energy levels. The latter is a cubic equation for $\omega^2$ and
can be solved for any field direction.


\begin{thebibliography}{11}

\bibitem{broholm-njp2007} M.B. Stone, C. Broholm, D.H. Reich, P. Schiffer,
O. Tchernyshyov, P. Vorderwisch and N.Harrison, New Journal of
Physics \textbf{9}, 31 (2007)

\bibitem{broholm-prb64} M.B. Stone, I. Zaliznyak, D.H.Reich and
C.Broholm Physical Review B \textbf{64}, 144405 (2001)

\bibitem{stone-nature} M.B. Stone, I.A. Zaliznyak, Tao Hong, C.L. Broholm and
D.H.Reich, Nature \textbf{440},187 (2006)

\bibitem{structure} L.P. Battaglia, A.B. Corradi, U. Geiser, R. Willett, A. Motori, F. Sandrolini, L. Antolini, T. Manfredini, L. Menaube
and G. Pellacani  J. Chem. Soc. Dalton. Trans. \textbf{2} 265 (1988)

\bibitem{Affleck} Ian Affleck, Physical Review B \textbf{46},
9002 (1992)

\bibitem{farmar}A.M. Farutin and V.I. Marchenko, Zh.Eksp.Teor.Fiz. \textbf{131} 860
(2007) (JETP \textbf{104} 751 (2007))

\bibitem{glazkov-tlcucl3}
V.N.Glazkov, A.I. Smirnov, H. Tanaka and A. Oosawa. Physical Review
B, \textbf{69} 184410 (2004).

\bibitem{glazkov-ntenp} V.N. Glazkov, A.I. Smirnov, A. Zheludev, and
B.C. Sales, Physical Review B \textbf{82}, 184406 (2010)

\bibitem{Abragam} A. Abragam, B. Bleaney, Electron paramagnetic resonance of transition ions.

\bibitem{tsvelik} A.M. Tsvelik, Physical Review  B \textbf{42}, 10499 (1990)

\bibitem{Kolezhuk} A.K. Kolezhuk, V.N. Glazkov, H. Tanaka, and A. Oosawa
Physical Review  B \textbf{70}, 020403 (2004)


\bibitem{stone-prl96} M.B. Stone, C. Broholm, D.H. Reich, O. Tchernyshyov, P. Vorderwisch and N. Harrison
Physical Review Letters \textbf{96}, 257203 (2006)

\bibitem{zaliznyak}L.-P. Regnault, I.A. Zaliznyak and S.V. Meshkov
J.Phys.:Condens.Matter \textbf{5} L677 (1993)

\end{thebibliography}
\end{document}